%% file: note_AF_-_v3.tex
\documentclass[reqno]{amsart}

\usepackage{amsmath,amssymb,amsfonts,amsthm}

\usepackage{mathtools}
\usepackage{xcolor}
\usepackage{mathpazo}
\usepackage{enumitem} 
\usepackage{graphicx}
\usepackage{subcaption}
\usepackage{booktabs}

\usepackage[backend=bibtex,style=alphabetic,hyperref, backref, backrefstyle=three, abbreviate=false, arxiv=pdf]{biblatex} 

\usepackage[linkcolor=blue,colorlinks=true, citecolor=blue, urlcolor=blue]{hyperref}

\usepackage[a4paper]{geometry}

\newtheorem{theorem}{Theorem}[section]

\newtheorem{proposition}[theorem]{Proposition}
\theoremstyle{definition}

\newtheorem{assumption}{Assumption}

\newtheorem{remarkx}[theorem]{Remark}
\newenvironment{remark}
{\pushQED{\qed}\begin{remarkx}}
	{\popQED\end{remarkx}}

\title{Note on a Theoretical Justification for Approximations of Arithmetic Forwards}
\author[\'Alvaro Romaniega]{\textbf{\'Alvaro Romaniega}\\
	\text{\footnotesize Internal Validation Quants - Risk Division}\\ Santander Group}
\date{\today}
\address{Internal Validation Quants - Risk Division - Santander Group}
\email{alvaroromaniega@gmail.com, alvaro.romaniega@gruposantander.com}
\thanks{I would like to thank César Romaniega for his thorough revision of the first version of this manuscript. I would also like to thank the IR internal validation team of Santander Group for the useful discussions that led to the writing of this manuscript.}

\begin{document}
	
	\maketitle

\begin{abstract}
	This note explores the theoretical justification for some approximations of arithmetic forwards ($F_a$) with weighted averages of overnight (ON) forwards ($F_k$). The central equation presented in this analysis is:
	\begin{equation*}
		F_a(0;T_s,T_e)=\frac{1}{\tau(T_s,T_e)}\sum_{k=1}^K \tau_k \mathcal{A}_k F_k\,,
	\end{equation*}
	with $\mathcal{A}_k$ being explicit model-dependent quantities, numerically stable and close to one under certain market scenarios. We will present computationally cheaper methods that approximate $F_a$, i.e., we will define some $\{\tilde{\mathcal{A}}_k\}_{k=1}^K$ such that
	\begin{equation*}
		F_a(0;T_s,T_e)\approx \frac{1}{\tau(T_s,T_e)}\sum_{k=1}^K \tau_k \tilde{\mathcal{A}}_k  F_k\,,
	\end{equation*}
	thereby gaining some intuition about the arithmetic factors $\mathcal{A}_k$. Additionally, theoretical bounds and closed-form expressions for the arithmetic factors $\mathcal{A}_k$ in the context of Gaussian HJM models are explored. Finally, we demonstrate that one of these forms can be closely aligned with an approximation suggested by Katsumi Takada in his work on the valuation of arithmetic averages of Fed Funds rates.
	\end{abstract}
\begingroup
\renewcommand{\addcontentsline}[3]{}
\section*{Disclaimer}
\textsc{\textit{The views and opinions expressed in this note are those of the author and do not necessarily reflect the official policy or position of} {Santander Group}}.
\endgroup
\tableofcontents

\tableofcontents

\section{Introduction}
In the realm of financial markets, various interest rate products are actively traded, such as interest rate swaps, basis swaps, and cross-currency swaps. Among these, the valuation of forward rates and their respective structures play a crucial role in pricing and risk management. One notable aspect of this valuation process involves the arithmetic average of overnight (ON) forward rates and its approximation.

Arithmetic averages of ON rates are particularly relevant in contexts where the Fed Funds (FF) rate is used, as seen in some swaps, liquid or OTC, with a floating leg based on FF or in Fed Fund Futures. While ON rates are often compounded daily in financial instruments like overnight index swaps, the arithmetic average of these rates offers a different perspective that requires careful consideration and adjustment for accurate valuation. This note aims to demonstrate the conditions under which arithmetic forwards can be closely approximated by a weighted average of the ON forwards. In particular, this aligns with an approximation given by some platforms, such as Murex, which provides the expression:
$$
F_a^\text{Murex}(0;T_s,T_e)\coloneqq \frac{1}{\tau(T_s,T_e)}\sum_{k=1}^K \tau_k F_k\,.
$$
Based on exact theoretical expressions for the arithmetic factors, we will provide a theoretical and numerical discussion under which market scenarios this approximation can be considered accurate and consider alternative and more accurate approximations. Additionally, analytical bounds and closed-form expressions for the arithmetic factors $\mathcal{A}_k$ in the context of Gaussian HJM models are explored. We also demonstrate that one of these forms can be closely aligned with an approximation suggested by Katsumi Takada in his work on the valuation of arithmetic averages of Fed Funds rates. See also \cite{Sko24} for a new approach in computing the convexity adjustment particularized for the SABR model.

\section{Notation}

In this note, we use the following notation:

\begin{itemize}[label={$\circ$}]
	\item $T_s, T_e$: The start and end dates of the interest period.
	\item $r$: The (collateralized) interest rate.
	\item $P(t, T)$: The price at time $t$ of a zero-coupon bond maturing at time $T$. $P(T) := P(0, T)$, i.e., the price at time 0 of a zero-coupon bond maturing at time $T$.
	\item $\tau_k$: The day count fraction, according to a given day-count convention, for the $k$-th period, $[T_{k-1}, T_k)$, within the interest period $\left[T_s, T_e\right]$. That is, it is the disjoint union,
	$$
	[T_s,T_e)=\bigsqcup_{k=0}^{K-1} [T_k, T_{k+1})
	$$
	\item $\tau(T_s, T_e)$: The day count fraction for the entire interest period $\left[T_s, T_e\right]$.
	\item $\mathbb{E}^{\cdot}$: The expectation operator, where the superscript denotes the measure under which the expectation is taken. Different measures used in this document include:
	\begin{itemize}
		\item $\mathbb{E}^Q$: The risk-neutral measure. Under this measure, the discounted price of a traded asset is a martingale. The numeraire is the money market account, denoted by $B$.
		\item $\mathbb{E}^{T_k}\coloneqq \mathbb{E}^{Q^{T_k}} $: The $T_k$-forward measure. Under this measure, the price of a zero-coupon bond maturing at time $T_k$ is used as the numeraire. 
		\item $\mathbb{E}^{T_e}\coloneqq \mathbb{E}^{Q^{T_e}}$: The $T_e$-forward measure. Under this measure, the price of a zero-coupon bond maturing at time $T_e$ is used as the numeraire. 
	\end{itemize}
	\item $\{R_k\}_{k=1,2,\ldots,K}$: The effective overnight rates fixed in the interest period $\left[T_s, T_e\right]$. In particular, for the period $[T_{k-1}, T_k)$.
	\item $R_g(T_s, T_e)$: The daily compounded ON rate over the interest period $\left[T_s, T_e\right]$, defined as:
	\[
	R_g(T_s, T_e) := \frac{1}{\tau(T_s, T_e)} \left( \prod_{k=1}^{K} (1 + \tau_k R_k) - 1 \right)\,.
	\]
	\item $R_a(T_s, T_e)$: The arithmetic average of ON rates (AAON) over the interest period $\left[T_s, T_e\right]$, defined as:
	\[
	R_a(T_s, T_e) := \frac{\sum_{k=1}^{K} \tau_k R_k}{\tau(T_s, T_e)}\,.
	\]
	\item Forward Contract: A forward contract based on a given rate $R$ is an agreement to exchange a specified amount of cash flow at a future date based on the interest rate $R$ determined over a certain period. If $V_t$ is the value of the floating leg at time $t$, we can define the forward at time $t$:
	\begin{equation*}
		F(t;T_s,T_e):=\frac{V_t}{\tau(T_s,T_e)P(T_e)}\,.
	\end{equation*}
	\item The (simply-compounded) forward rate, at time $t=0$, associated with the $k$-th period within the interest interval $\left[T_s, T_e\right]$, which can be defined as:
	$$
	F_k\coloneqq \frac{1}{\tau_k}\left(\frac{P(T_{k-1})}{P(T_{k})}-1\right)\,.
	$$
\end{itemize}

	\section{Unweighted approximation and closed form for the arithmetic factors}
	\subsection{Main idea}
	The present value of the floating leg at time $t=0$ with an AAON over the interest period $\left[T_s, T_e\right]$ with unit nominal amount is given by
	\begin{align*}
		V_0^a & =\mathbb{E}^Q\left(e^{-\int_0^{T_e} r(u) d u} \tau\left(T_s, T_e\right) R_a\right) \\
		& =\mathbb{E}^Q\left(e^{-\int_0^{T_e} r(u) d u}\sum_{k=1}^K \tau_k R_k\right)\,.
	\end{align*}
	By linearity,
	\begin{equation*}
		V_0^a = \sum_{k=1}^K \tau_k \mathbb{E}^Q\left(e^{-\int_0^{T_e} r(u) d u} R_k\right)\,.
	\end{equation*}
	Given that $r \ge 0$ and $T_k \le T_e$, there exist $\{A_k\}_{k=1}^K$ with $A_k \le 1$, see Remark \ref{rem:Ak_leq_1}, such that
	\begin{equation*}
		\sum_{k=1}^K \tau_k \mathbb{E}^Q\left(e^{-\int_0^{T_e} r(u) d u} R_k\right)=\sum_{k=1}^K \tau_k A_k \mathbb{E}^Q\left(e^{-\int_0^{T_k} r(u) d u} R_k\right)\,,
	\end{equation*}
	with, heuristically speaking,
	\begin{equation*}
		A_k\approx  \frac{\mathbb{E}^Q\left(e^{-\int_{0}^{T_e} r(u) d u}\right)}{\mathbb{E}^Q\left(e^{-\int_{0}^{T_k} r(u) d u}\right)}=\frac{P(T_e)}{P(T_k)}\,,
	\end{equation*}
	see Section \ref{sec:justification} for further justification. We can define the numeraire $N_t^k:= P(t, T_k)$ and perform multiple changes of measure such that
	\begin{equation*}
		\sum_{k=1}^K \tau_k A_k \mathbb{E}^Q\left(e^{-\int_0^{T_k} r(u) d u} R_k\right)=\sum_{k=1}^K \tau_k A_k P(T_k) \mathbb{E}^{T_k}\left(R_k\right)\,.
	\end{equation*}
	By standard results, see, for instance, Lemma 4.2.3 in \cite{AP10a} (or (146) of \cite{AB13} and above (6.47) in \cite{AP10a}) for the multicurve case),
	\begin{equation}\label{eq:fwds}
		\mathbb{E}^{T_k}\left(R_k\right)=F_k\,,
	\end{equation}
	the forward associated with the date $T_k$. All in all,
	\begin{equation}\label{eq:fwd_value}
		V_0^a = \sum_{k=1}^K \tau_k A_k P(T_k) F_k=P(T_e)\sum_{k=1}^K \tau_k \mathcal{A}_k F_k\,,
	\end{equation}
	where we have define our arithmetic factors as
	\begin{equation*}
		\mathcal{A}_k:=A_k \frac{P(T_k)}{P(T_e)}\,.
	\end{equation*}
	These are model-dependent quantities that under the aforementioned approximation are close to one. Indeed, as mentioned above,
	\begin{equation}\label{eq:coef approx}
		\mathcal{A}_k\approx \frac{P(T_e)}{P(T_k)}\frac{P(T_k)}{P(T_e)}=1\,.
	\end{equation}
	Thus, see also Equation (12) in \cite{Tak11} for a slightly different definition, by definition and \eqref{eq:fwd_value},
	\begin{equation*}
		F_a(0;T_s,T_e):=\frac{V_0^a}{\tau(T_s,T_e)P(T_e)}=\frac{1}{\tau(T_s,T_e)}\sum_{k=1}^K \tau_k \mathcal{A}_k F_k\,.
	\end{equation*}
	Using the approximation of \eqref{eq:coef approx} then
	\begin{equation*}
		F_a(0;T_s,T_e)\approx \frac{1}{\tau(T_s,T_e)}\sum_{k=1}^K \tau_k F_k\,.
	\end{equation*}
	\subsection{Rigorous proof and closed form of the arithmetic factor $\mathcal{A}_k$}
	\label{sec:justification}
	By definition, with $B(t):=e^{\int_0^t r}$ the money market account with the collateralized rate,
	\begin{equation}\label{eq:def_A_k}
		A_{k} = \frac{\mathbb{E}^{Q} \left( R_k B^{-1}(T_{e}) \right)}{\mathbb{E}^{Q} \left( R_k B^{-1}(T_{k}) \right)} = \frac{\mathbb{E}^{T_e} \left( R_k  \right) P(T_e)}{\mathbb{E}^{T_k} \left( R_k \right) P(T_k)}\rightarrow \mathcal{A}_{k}=\frac{\mathbb{E}^{T_e} \left( R_k  \right)}{\mathbb{E}^{T_k} \left( R_k \right)} \,,
	\end{equation}
	where we have used the change of numeraire formula twice. We know\footnote{We are considering the simpler case, $t=0$, where $\mathcal{F}_0$ is the trivial $\sigma$-algebra. For the case $t>0$, we would use Abstract Bayes' Theorem, for instance, Lemma 19.7 in \cite{Sch21}, Lemma 8.6.2 in \cite{Oks13}, Appendix C.3 of \cite{Bjo20} or page 9 of \cite{AP10a}, for measures $P,Q$, with Radon-Nikodym derivative $L$ and a $\mathcal{F}_T$-measurable function $X$,
		\begin{equation}\label{eq:Bayes_th}
			\mathbb{E}^{Q} \left( {X} | \mathcal{F}_{t} \right) = \mathbb{E}^{P} \left( X \frac{L_{T}}{L_{t}} \bigg| \mathcal{F}_{t} \right)\,.
	\end{equation}}, for instance, Chapter 15 of \cite{Bjo20}, Theorem 1.4.2 of \cite{AP10a}, that the likelihood process, for a given filtration $\{\mathcal{F}_t\}_t$, is
	\begin{equation*}
		L_{t}^{k} := \frac{dQ^{T_k}}{dQ^{T_e}} \bigg|_{\mathcal{F}_{t}} = \frac{P(t,T_k)/P(T_k)}{P(t,T_e)/P(T_e)}\,,
	\end{equation*}
	so we arrive at the expression
	\begin{equation*}
		\mathbb{E}^{{T_k}} \left( R_k \right) = \mathbb{E}^{{T_e}} \left( R_k \cdot \frac{P(T_{e})} {P(T_{k}, T_{e})P(T_{k})} \right)\,. 
	\end{equation*}
	Therefore,
	\begin{equation*}
		\mathcal{A}_{k} = \mathbb{E}^{{T_e}} \left( \frac{R_k}{\mathbb{E}^{{T_e}}(R_k)} \cdot \frac{P(T_{e})}{P(T_{k}, T_{e}) \cdot P(T_{k})} \right)^{-1}\,,
	\end{equation*}
	where the economic interpretation of the second term corresponds to the strategy of rolling the bond versus not rolling it at $T_k$.
	Using the martingale pricing formula for a given numeraire\footnote{Also, as $\frac{P(t,T_k)}{P(t, T_{e})}$ is a $T_e$-martingale,
	$$
	\mathbb{E}^{T_e} \left( \frac{1}{P(T_{k}, T_{e})} \right) = \mathbb{E}^{T_e} \left( \frac{P(T_k,T_k)}{P(T_{k}, T_{e})} \right) = \frac{P(T_k)}{P(T_{e})}\,.
	$$
	}
	\begin{equation}
		\mathbb{E}^{T_e} \left( \frac{1}{P(T_{k}, T_{e})} \right) = \mathbb{E}^{Q} \left( \frac{1}{B(T_{k})} \right) \frac1{P(0, T_{e})} = \frac{P(T_k)}{P(T_{e})}\,.
	\end{equation}
	Thus, finally,
\begin{equation}\label{eq:final Ak}
	\mathcal{A}_k = \mathbb{E}^{T_{e}} \left( \frac{R_{k}}{\mathbb{E}^{T_{e}}(R_{k})} \cdot \frac{P(T_{k}, T_{e})^{-1}}{\mathbb{E}^{T_{e}}(P(T_{k}, T_{e})^{-1})} \right)^{-1}\,.
\end{equation}
	Intuitively, this expression is close to one as it represents the expectation of two random variables that each have an expectation of one. Trivial sufficient conditions for the approximation to hold are:
	\begin{enumerate}
		\item If $T_k$ is close to $T_e$, then $P(T_k, T_e) \approx 1$. 
		Therefore,
		$$
		\mathcal{A}_{k} \approx \mathbb{E}^{{T_e}} \left( \frac{R_k}{\mathbb{E}^{{T_e}}(R_k)}\cdot 1 \right)^{-1}=1\,.
		$$
		\item If interest rate volatility is low for the $[T_{k-1}, T_{k}]$-forward, so, $R_k\approx \mathbb{E}^{T_{e}}(R_{k})$. Thus,
		$$
			\mathcal{A}_{k} \approx \mathbb{E}^{{T_e}} \left( 1\cdot\frac{P(T_{k}, T_{e})^{-1}}{\mathbb{E}^{T_{e}}(P(T_{k}, T_{e})^{-1})} \right)^{-1}=1\,.
		$$
		\item Obviously, for deterministic rates the quantity is exactly one.
	\end{enumerate}
\subsubsection{Alternative expressions}\label{sec:alt_exp}
Similarly, we could have chosen $T=T_{k-1}$ in \eqref{eq:Bayes_th} or use the tower property in \eqref{eq:final Ak} with $\mathcal{F}_{T_{k-1}}$ to arrive at the equivalent expression
\begin{equation}\label{eq:final Ak_alt_mean}
		\mathcal{A}_{k}=\mathbb{E}^{{T_e}} \left( \frac{R_k}{\mathbb{E}^{T_{e}}(R_{k})} \cdot \frac{P(T_{k-1},T_k)P(T_{e})} {P(T_{k-1}, T_{e})P(T_{k})} \right)^{-1}=\mathbb{E}^{{T_e}} \left( \frac{R_k}{\mathbb{E}^{T_{e}}(R_{k})}\frac{P(T_{k-1},T_k,T_{e})^{-1}}{\mathbb{E}^{{T_e}}(P(T_{k-1},T_k,T_{e})^{-1})} \right)^{-1}\,, 
\end{equation}
as $P(t,T_{k},T_e)^{-1}$ is a $T_e$-martingale. More compactly,
\begin{equation}\label{eq:final Ak_means}
	\mathcal{A}_{k}=\mathbb{E}^{{T_e}} \left( \bar{R}_k^{T_e} \cdot \overline{P^{-1}}^{T_e}(T_{k-1},T_k,T_{e})\right)^{-1}=\mathbb{E}^{{T_e}} \left( \bar{R}^{T_e}_{k} \cdot \overline{P^{-1}}^{T_e}(T_{k}, T_{e})^{-1}\right)^{-1}\,, 
\end{equation}
where  $\bar{X}^{T_e}\coloneqq X/\mathbb{E}^{T_e}(X)$. Similarly, as the expectation of $R_k$ under $T_k$ is known, $F_k$, but not under $T_e$, we could use the inverse change of measure
\begin{equation*}
		\tilde L_{t}^{k} := \frac{dQ^{T_e}}{dQ^{T_k}} \bigg|_{\mathcal{F}_{t}} = \frac{P(t,T_e)/P(T_e)}{P(t,T_k)/P(T_k)}=\left(L_t^k\right)^{-1}\,.
\end{equation*}
In the same vein,
\begin{equation}\label{eq:final Ak_alt_Tk}
	\mathcal{A}_{k}=\mathbb{E}^{{T_k}} \left( \frac{{R}_k}{\mathbb{E}^{T_{k}}(R_{k})} \cdot \frac{{P}(T_{k-1}, T_e) P(T_{k-1})}{P(T_{k-1}, T_k)P(T_e)} \right)=\mathbb{E}^{{T_k}} \left( \frac{{R}_k}{\mathbb{E}^{T_{k}}(R_{k})}\frac{{P}(T_{k},T_e)P(T_k)}{P(T_e)}\right)\,,
\end{equation}
Also,
\begin{equation}\label{eq:final Ak_alt_Tk_mean}
	\mathcal{A}_{k}=\mathbb{E}^{{T_k}} \left( \bar{R}^{T_k}_k \cdot \bar{P}^{T_k}(T_{k-1}, T_k, T_e) \right)=\mathbb{E}^{{T_k}} \left( \bar{R}^{T_k}_k\bar{P}^{T_k}(T_{k},T_e)\right)\,, 
\end{equation}
where we have used, as above, the definition $\bar{X}^{T_k}\coloneqq X/\mathbb{E}^{T_k}(X)$, the fact that $P(T_{k-1}, T_k, T_e)$ is a $T_k$ martingale and 
\begin{equation*}
\mathbb{E}^{T_k} \left({P(T_{k}, T_{e})} \right) = \mathbb{E}^{Q} \left( \frac{P(T_k,T_e)}{B(T_{k})} \right) \frac1{P(0, T_{k})} = \frac{P(T_e)}{P(T_{k})}\,,
\end{equation*}
by a standard application of the tower property. Note that all the expressions share the same structure of the product of two ``normalized'' random variables. The differences lie in the:
\begin{itemize}
	\item \textit{Measurability}: \( R_k \), \( P(T_{k-1}, T) \) for any \( T \ge T_{k-1} \) are \( \mathcal{F}_{T_{k-1}} \)-measurable, but \( P(T_{k}, T_e) \) is \( \mathcal{F}_{T_{k}} \)-measurable.
		\item \textit{Terminal measure}: The first expressions use a common measure for the calculations, $Q^{T_e}$, but the latter use a different measure for each factor, $Q^{T_k}$. This is advantageous in general, but it also implies not using \eqref{eq:fwds}.
		\item \textit{Variance reduction}: We can also compare these expressions with the definition of $\mathcal{A}_k$ in \eqref{eq:def_A_k}. Although these expressions for computing the arithmetic factors appear more complex due to their incorporation of the Radon-Nikodym derivative, it can be particularly useful in practice because it yields more stable and accurate results in Monte Carlo simulations, mitigating the impact of estimation errors. We will explore this in more detail in the following sections.
\end{itemize}

\section{Some intuition and example of numerical computation}
\begin{figure}[h!]
	\centering
	\begin{subfigure}[b]{0.925\textwidth}
		\centering
		\includegraphics[width=\linewidth]{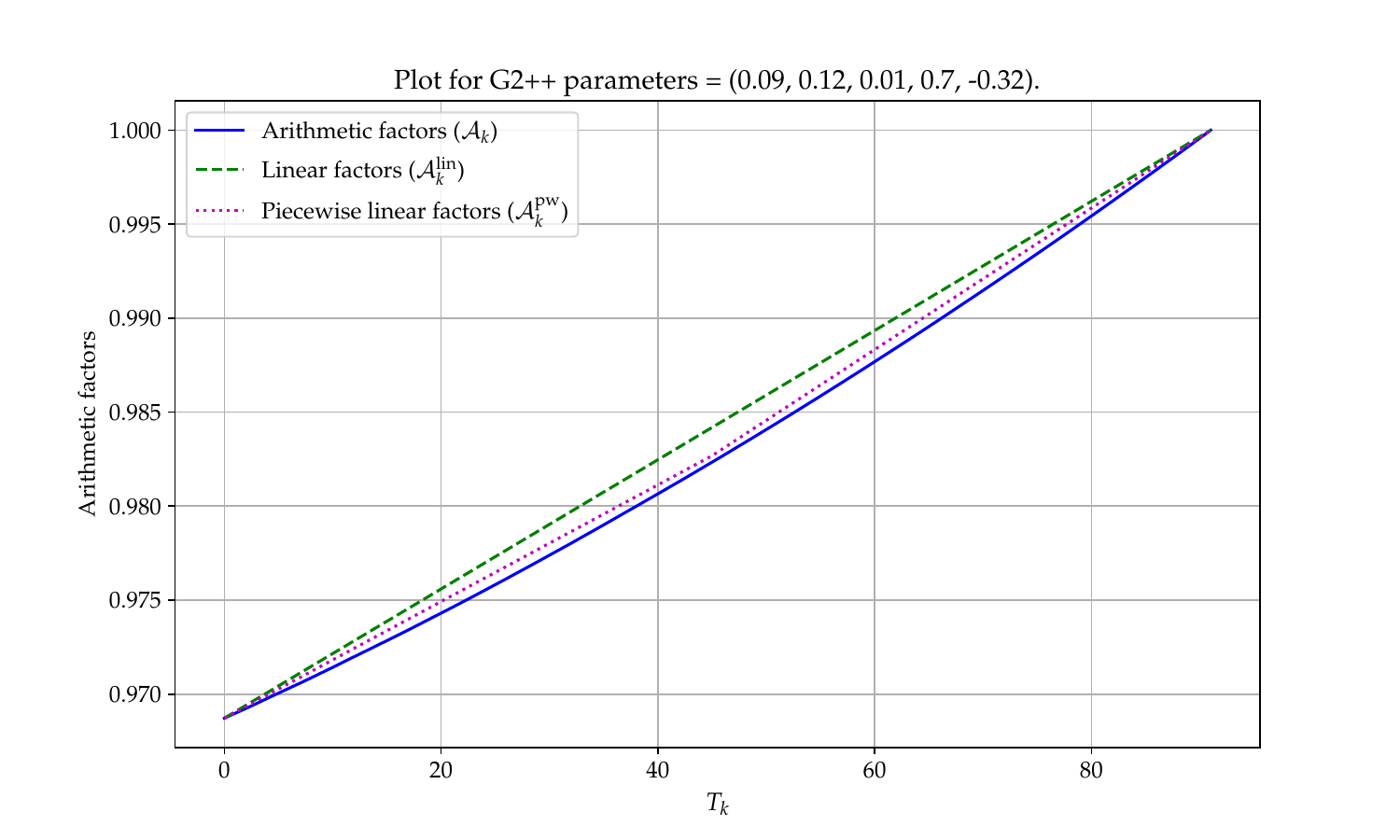}
		\caption{$T_e-T_s=$3 months.}
		\label{fig:plot_tenor_3M}
	\end{subfigure}
	\hfill
	\begin{subfigure}[b]{0.925\textwidth}
		\centering
		\includegraphics[width=\linewidth]{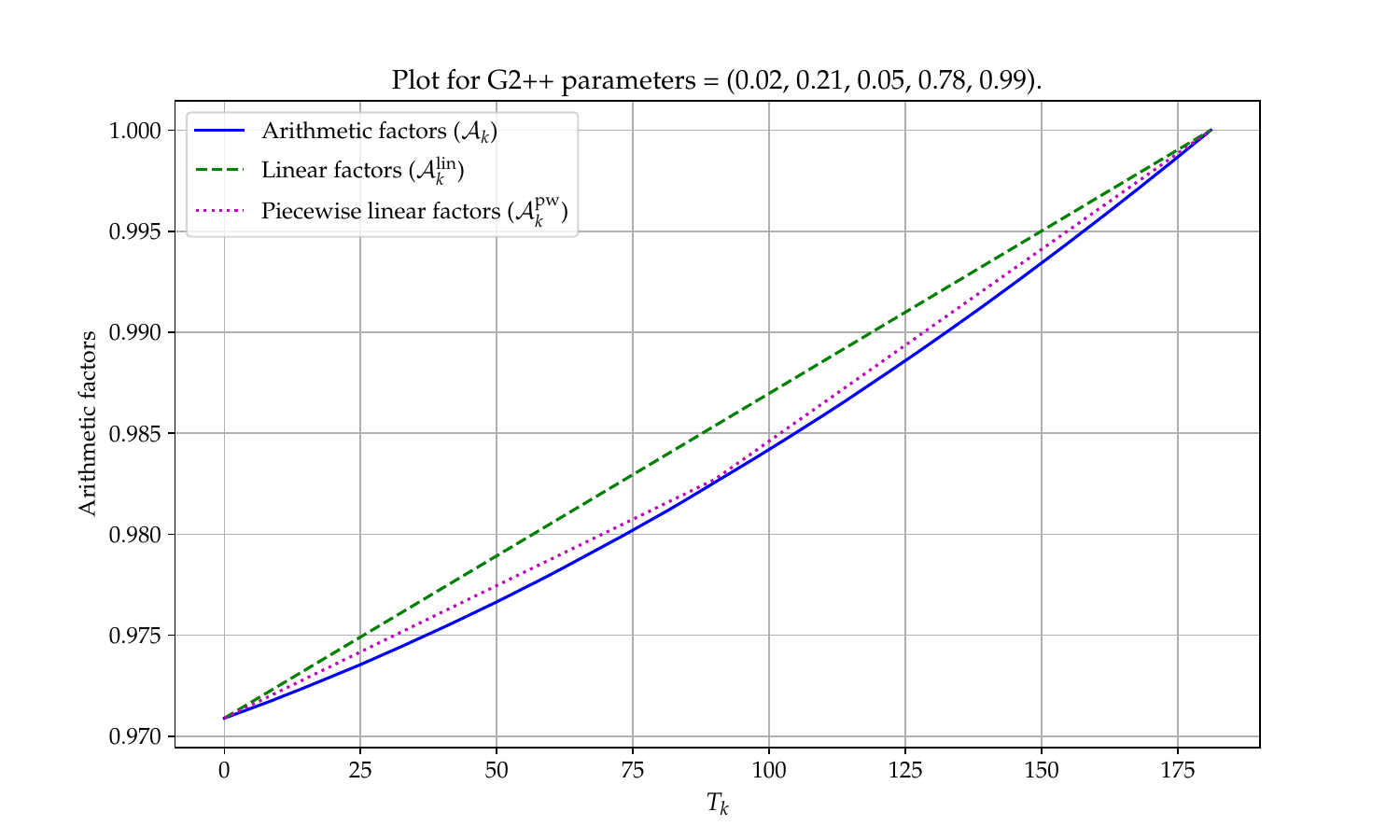}
		\caption{$T_e-T_s=$6 months.}
		\label{fig:plot_tenor_6M}
	\end{subfigure}
	\caption{Plot of $(\mathcal{A}_k)_{k=1}^K$ in blue, $(\mathcal{A}^\text{lin}_k)_{k=1}^K$ in green, and $(\mathcal{A}^\text{pw}_k)_{k=1}^K$ in pink. $T_s = 12$ months. $\mathcal{A}_k$ was computed using Monte Carlo with $10^6$ simulations.}
	
	\label{fig:combined}
\end{figure}

Although the arithmetic factors are generally close to one, this approximation can be far from accurate if, for instance, interest rate volatility is high and we are far from the end date. For the sake of conciseness, assume that the evolution of the instantaneous short-rate process under the risk-neutral measure \( Q \) is given by a Markovian\footnote{Understood in the sense of Proposition 12.1.1 of \cite{AP10b}. There, as the instantaneous forward rate is $f(t,T)=f(0, T)+\Omega(t, T)+h(T)^{\top} z(t)$, i.e., a function of the state variables $z$ at $t$ and some deterministic functions, then
	$$
	P(t, T)=e^{-\int_t^T f(t, u) d u}=e^{\int_t^T(f(0, u)+\Omega(t, u))du+\left(\int_t^{T} h(u)^{\top}du\right) \cdot z(t)}=P(t, T, z(t))\,.
	$$ 
} \( n \)-factor model:

\begin{equation}\label{eq:sr model}
	r(t) = \sum_{i=1}^n x_i(t) + \varphi(t)=\sum_{i=0}^n x_i(t), \quad r(0) = r_0,
\end{equation}
where \(x_0(t)\coloneqq\varphi(t)\) and the Itô processes \(\{x_i(t) : t \geq 0\}_{i=1}^n\) satisfy the following stochastic differential equations (Ornstein–Uhlenbeck process):
$$
	\begin{cases*}
		dx_i(t) = -a_i x_i(t) dt + \sigma_i dW_i(t), \\
		x_i(0) = 0,
	\end{cases*}
	 \quad i = 1, \ldots, n\,.
$$
Here, $\left(W_i(t)\right)_{i=1}^n$ is a \( n \)-dimensional Brownian motion with instantaneous correlations given by the quadratic covariation:
\begin{equation}
	\langle W_i(t), W_j(t)\rangle = \rho_{ij}t, \quad \text{for} \; i, j = 1, \ldots, n,
\end{equation}
where \( r_0, a_i, \sigma_i \) are positive constants, \(\rho_{ij}\) are the correlation coefficients with \(-1 \leq \rho_{ij} \leq 1\), and \(\varphi(t)\) is a deterministic function that recovers the spot discount curve given by the market, $P^M(0, T)$. We denote by \(\mathcal{F}_t\) the $\sigma$-field generated\footnote{More precisely, see Definition 9.19 of \cite{Dri19},
	\begin{equation*}
		\mathcal{F}_t := \sigma\left(x(s), s \leq t\right) := \sigma \left( \bigcup_{s \leq t} x(s)^{-1}(\mathcal{B}(\mathbb{R}^n)) \right) = \text{minimal } \sigma\text{-algebra such that all } x(s) \text{ are measurable.}
\end{equation*}} by the processes \(x(t)\coloneqq\left(x_i(t)\right)_{i=1}^n \) up to time \( t \).
\begin{remark}
The procedure could be generalized to other Markovian models (with something similar to equation \eqref{eq:markovian_rep}); however, we have chosen this model for simplicity to illustrate the procedure and to provide some properties and intuition, although we will try to keep the exposition as general as possible. Furthermore, in the next section, we will investigate the analytical approach. A further and obvious line of research would be to extend the results to other models.
\end{remark}
Therefore,
\begin{align*}
	P(t, T) &=\mathbb{E}^Q \left( \exp \left( -\int_t^T r_s ds \right) \bigg| \mathcal{F}_t \right) \quad \text{(martingale measure)} \\
	&= \mathbb{E}^Q \left( \exp \left( -\int_t^T r_s ds \right) \bigg| \sigma\left(x_t^1, x_t^2, \ldots, x_t^n\right) \right) \quad \text{(Markovian property)} \\
	&= \mathbb{E}^Q \left( \exp \left( -\int_t^T \left( \sum_{i=0}^n x_s^i \right) ds \right) \bigg| \sigma\left(x_t^1, x_t^2, \ldots, x_t^n\right) \right) \quad \text{(definition of $r_s$)} \\
	&= H_t^T(x_t^1, x_t^2, \ldots, x_t^n)\,, \quad \text{(Factorization Lemma)}
\end{align*}
where $x_t = (x_t^1, x_t^2, \ldots, x_t^n)$ and see \cite[Lemma 9.42]{Dri19} for the Factorization Lemma (Doob–Dynkin Lemma), which guarantees the existence\footnote{This step is usually avoided unconsciously, cf. the law of the unconscious statistician, by making a notation convention for a random variable $X$, $\mathbb{E}(F|\sigma(X))=:\mathbb{E}(F|X)$, a mathematical statement, $\exists~H$ measurable such that $\mathbb{E}(F|\sigma(X))=H(X)$. Actually, for a particular value $x$, $\mathbb{E}(F|X=x)\coloneqq H(x)$; see, for instance, the discussion below Lemma 1.2 in \cite{Sha03}. } of the measurable function $H_t^T$. Thus, using that
\[
R_k = \frac{1}{\tau_k} \left( \frac{1}{P(T_{k-1}, T_{k})} - 1 \right)
\]
and
\begin{equation}
	P(T_{k-1}, T_{k}) = H_{T_{k-1}}^{T_{k}}(x_{T_{k-1}})\,, \, P(T_{k-1}, T_{e}) = H_{T_{k-1}}^{T_{e}}(x_{T_{k-1}})\,,\, P(T_k, T_e) = H_{T_k}^{T_e}(x_{T_k})\,,
	\label{eq:markovian_rep}
\end{equation}
we could compute the arithmetic factors $\mathcal{A}_k$ in \eqref{eq:final Ak}, \eqref{eq:final Ak_means}, \eqref{eq:final Ak_alt_Tk_mean}
as we know the distribution, or an approximation, of the random vector $x_{T_{k-1}}$ so it is a simple expectation.
Hereafter, for the sake of simplicity and clarity, consider the G2++ model, that is, just two factors $(x_t, y_t)$ \cite[Chapter 4]{BM06}. We know that\footnote{For a general multi-factor Gaussian model see Corollary 12.1.3 of \cite{AP10b}. See also Section \ref{sec:GHJM}.}, \cite[Corollary 4.2.1]{BM06},
\begin{equation}
	\begin{aligned}
		V(t,T) &= \frac{\sigma^2}{a^2} \left[ T - t + \frac{2}{a} e^{-a(T-t)} - \frac{1}{2a} e^{-2a(T-t)} - \frac{3}{2a} \right] \\
		&\quad + \frac{\eta^2}{b^2} \left[ T - t + \frac{2}{b} e^{-b(T-t)} - \frac{1}{2b} e^{-2b(T-t)} - \frac{3}{2b} \right] \\
		&\quad + 2 \rho \frac{\sigma \eta}{ab} \left[ T - t + \frac{e^{-a(T-t)} - 1}{a} + \frac{e^{-b(T-t)} - 1}{b} - \frac{e^{-(a+b)(T-t)} - 1}{a+b} \right]
	\end{aligned}
\end{equation}
and
\begin{equation}
	P(t,T, x_t, y_t) = \frac{P^M(0,T)}{P^M(0,t)} \exp \left\{ A(t,T, x_t, y_t) \right\}\,,
\end{equation}
where
\begin{equation}
	A(t,T, x_t, y_t) := \frac{1}{2} \left[ V(t,T) - V(0,T) + V(0,t) \right] - \left[ \frac{1 - e^{-a(T-t)}}{a} x(t) + \frac{1 - e^{-b(T-t)}}{b} y(t) \right]\,,
\end{equation}
and $P^M$ market bonds. As G2++ fits the spot discount bonds, $P^M$, then
\begin{equation*}
\frac{P(T_{e})}{P(T_{k}, T_{e}) \cdot P(T_{k})}=\exp \left\{ -A(T_k,T_e, x_{T_k}, y_{T_k}) \right\}\,.
\end{equation*}
Since $V(t,T)$ is a function depending only on the tenor, i.e., $V(t,T)=\tilde V(T-t)$ with $\tilde V(0)=0$. By Taylor's Theorem, $\tilde V(T)=\tilde V(t)+O(T-t)$ with the constant depending only on the model parameters. Therefore, we can arrive at:
\begin{align*}
	\frac{P(T_k, T_e)\cdot P(T_k)}{P(T_e)} &=1 + O(T_e-T_k) - (x_{T_k}+y_{T_k})(T_e-T_k) \\ &-\frac12(a\cdot x_{T_k}^2+b\cdot y_{T_k}^2)(T_e-T_k)^2 + o_{x,y}\left((T_e-T_k)^2\right)\,,
\end{align*}
where the little-o term depends on the stochastic factors at time $T_k$, not just the model parameters.
Thus, heuristically, given this expression, \eqref{eq:final Ak}, and the fact that if $T_K=T_e$, $\mathcal{A}_k=1$, we can propose the simple linear approximation:
$$
\mathcal{A}_k^\text{lin}\coloneqq \mathcal{A}_1 + (T_k-T_1)\frac{1-\mathcal{A}_1}{T_e-T_1}\,.
$$
The obvious advantage of this formulation is that, while taking into account the ``convexity'' adjustment of arithmetic forwards, we just need to compute the expectation in \eqref{eq:def_A_k}, not for every $k$, reducing the computational cost or increasing the analytical tractability. From the expression above, we can see that the quadratic term of the stochastic factors and, in general, the convexity of the bond price function, will generally make this approximation an upper bound, cf. Proposition \ref{prop:Ak_le_1}. Obviously, we can refine this approximation by taking a midpoint and performing a piecewise linear interpolation, so we can account for the curvature of the arithmetic factors with respect to the tenor, see Figure \ref{fig:combined}. That is,

\[
\mathcal{A}_k^\text{pw} =
\begin{cases} 
	\mathcal{A}_1 + (T_k - T_1) \frac{\mathcal{A}_m - \mathcal{A}_1}{T_m - T_1}, & \text{for } k \leq m\,, \\
	\mathcal{A}_m + (T_k - T_m) \frac{1 - \mathcal{A}_m}{T_e - T_m}, & \text{for } k > m \,,	
\end{cases}
\]
where \(T_m\) is the time at the midpoint.

\subsection{Numerical simulations}
For the sake of simplicity, take a spot discount curve with a constant discount rate of $r=0.05$ (so all the forwards are approximately the same number) and some random parameters.
\begin{table}[!ht]
	\centering
\input{tables/results_tenor_3M_delay_1M_no_sims_100e3_models_False.tex}
\caption{Results for $T_s = 1$ month and $T_e - T_s = 3$ months. $F_a$ was computed using Monte Carlo with $10^5$ simulations. The linear approximation reduces the error by a factor of 2, and the piecewise linear approximation reduces the error by a factor of 4 with respect to the linear one.}
\end{table}

\begin{table}[!ht]
	\centering
\input{tables/results_tenor_6M_delay_12M_no_sims_100e3_models_False.tex}
	\caption{Results for $T_s = 12$ months and $T_e - T_s = 6$ months. $F_a$ was computed using Monte Carlo with $10^5$ simulations. The linear approximation reduces the error by a factor of more than 8, and the piecewise linear approximation reduces the error by a factor of slightly less than 4 with respect to the linear one.}
	
\end{table}
 
\begin{remark}\label{rem:U_shape}
	If $T_s$ is close to zero, then the volatility of $r(T_1)$ is low, see \cite[(4.19)]{BM06}, so point (2) of the previous section applies, i.e.,
	$$
	\mathcal{A}_{1} \approx \mathbb{E}^{{T_e}} \left( 1\cdot\frac{P(T_{1}, T_{e})^{-1}}{\mathbb{E}^{T_{e}}(P(T_{1}, T_{e})^{-1})} \right)^{-1}=1\,.
	$$
	$\mathcal{A}_1$ is also close to one, as is $\mathcal{A}_K$, which is close to one because of item (1) of the previous section. Thus, the plot of $\mathcal{A}_k$ can present a U-shaped curve, see Figure \ref{fig:HW_A_k}. Thus, using a mid-point makes more sense.
\end{remark}
\subsection{Alternative expressions and numerical performance}
\begin{figure}[h!]
	\centering
	\begin{subfigure}[b]{0.75\textwidth}
		\centering
		\includegraphics[width=\linewidth]{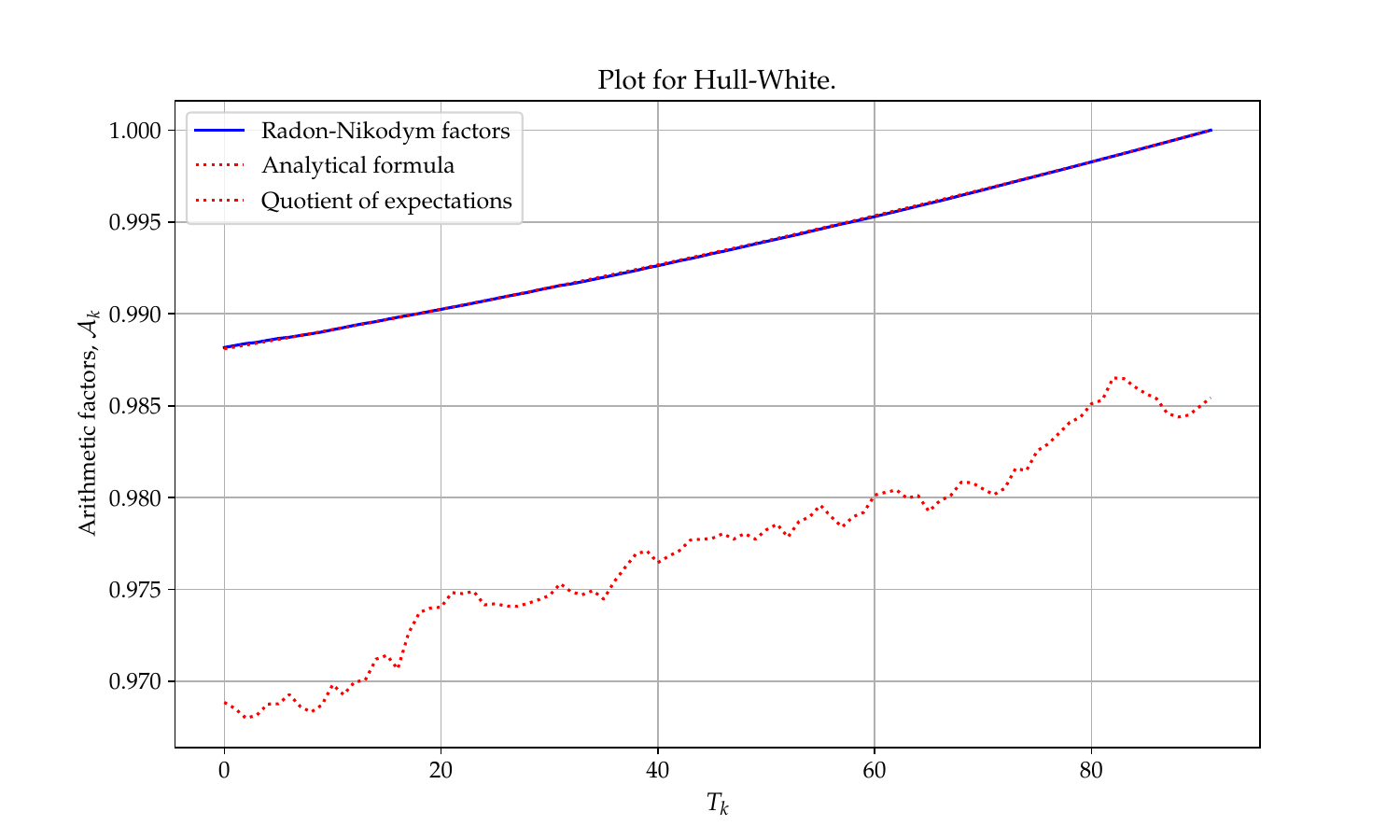}
		\caption{$10^4$ simulations.}
		\label{fig:comparison_10e3}
	\end{subfigure}
	\hfill
	\begin{subfigure}[b]{0.75\textwidth}
		\centering
		\includegraphics[width=\linewidth]{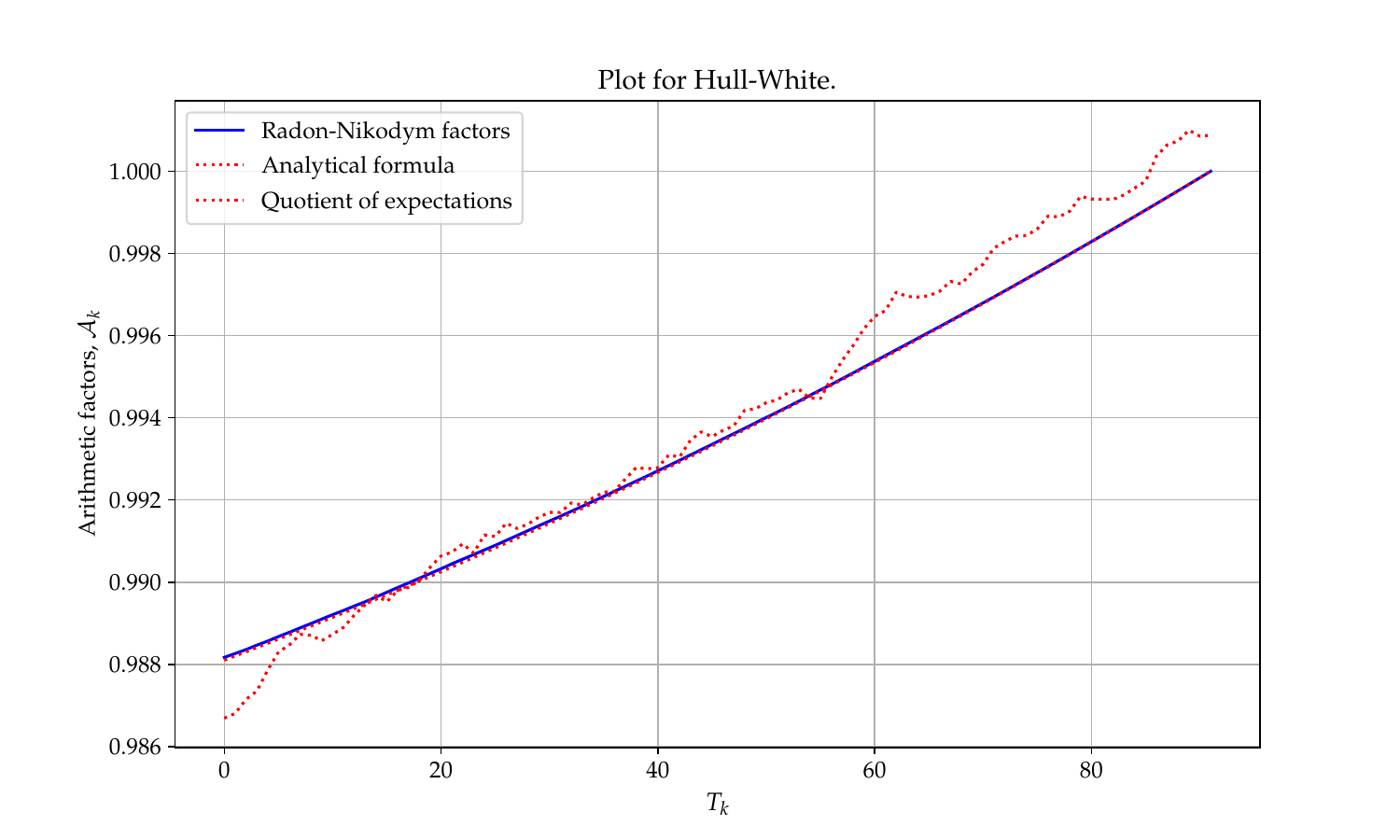}
		\caption{$10^5$ simulations.}
		\label{fig:comparison_100e3}
	\end{subfigure}
	\hfill
	\begin{subfigure}[b]{0.75\textwidth}
		\centering
		\includegraphics[width=\linewidth]{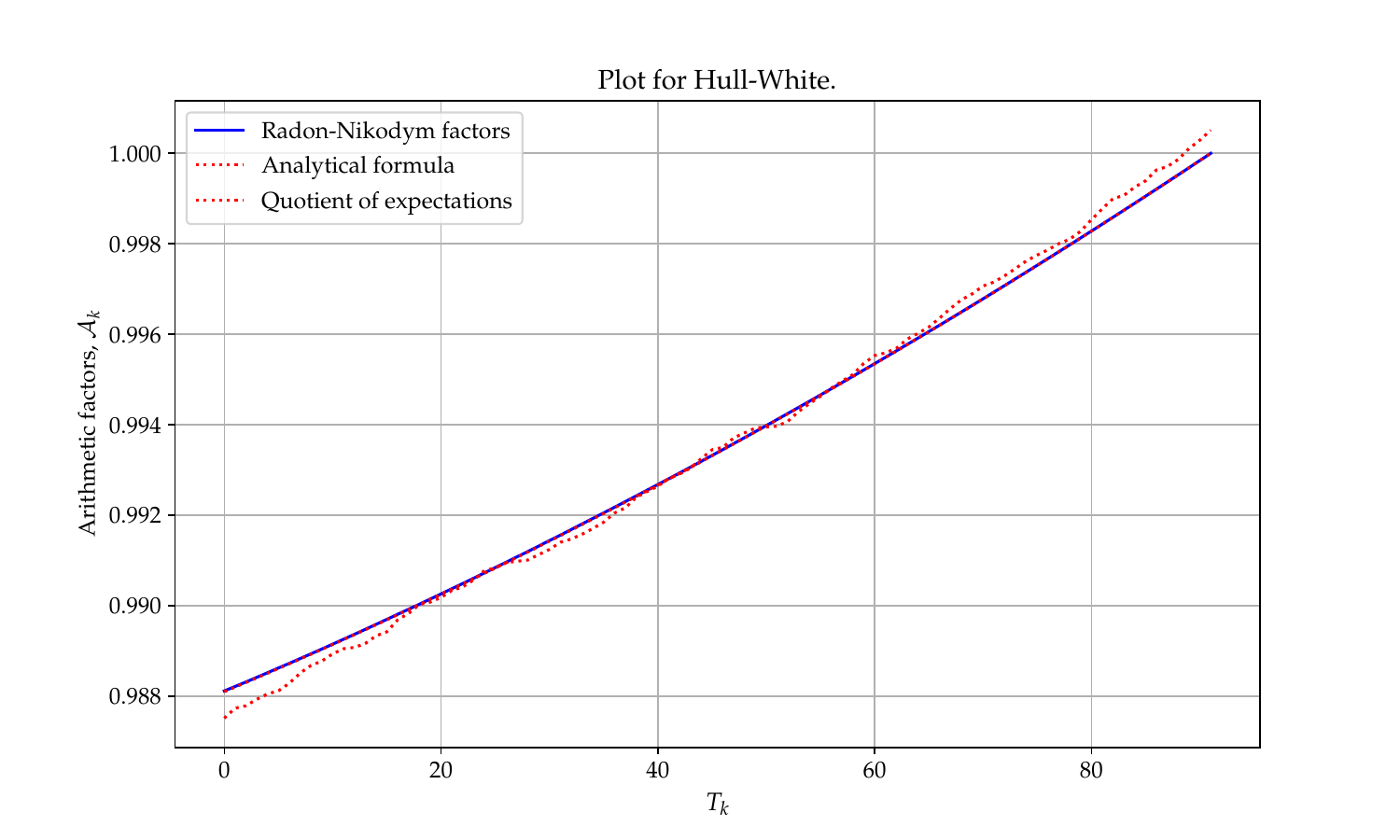}
		\caption{$10^6$ simulations.}
		\label{fig:comparison_1000e3}
	\end{subfigure}
	\caption{Comparison of the different expressions for the arithmetic factors. The number of simulations is varied in each plot. Here, $T_s=12$ months and $T_e-T_s=3$ months.}
	\label{fig:comparison}
\end{figure}

As we mentioned before, there are several alternative expressions for the arithmetic factors. We can compare the numerical performance of the different expressions using the same set of parameters and number of simulations. This is done in Figure \ref{fig:comparison}, where we have plot the arithmetic factors using \eqref{eq:def_A_k} and Section \ref{sec:alt_exp}, but varying the number of simulations.

As we can see, expression \eqref{eq:def_A_k} is less accurate than \eqref{eq:final Ak_alt_mean}, the one we have used for simulations. The latter aligns with the analytical results that are available for the Hull-White model with constant parameters, see
 Section \ref{sec:HW_case}. This model corresponds to the case of $n=1$ in \eqref{eq:sr model}. As we can see, the expression with the product of two normalized variables performs better and needs less simulations to converge to the true value, reducing the computational cost.

\begin{figure}[h!]
	\centering
	\begin{subfigure}[b]{0.9\textwidth}
		\centering
		\includegraphics[width=\linewidth]{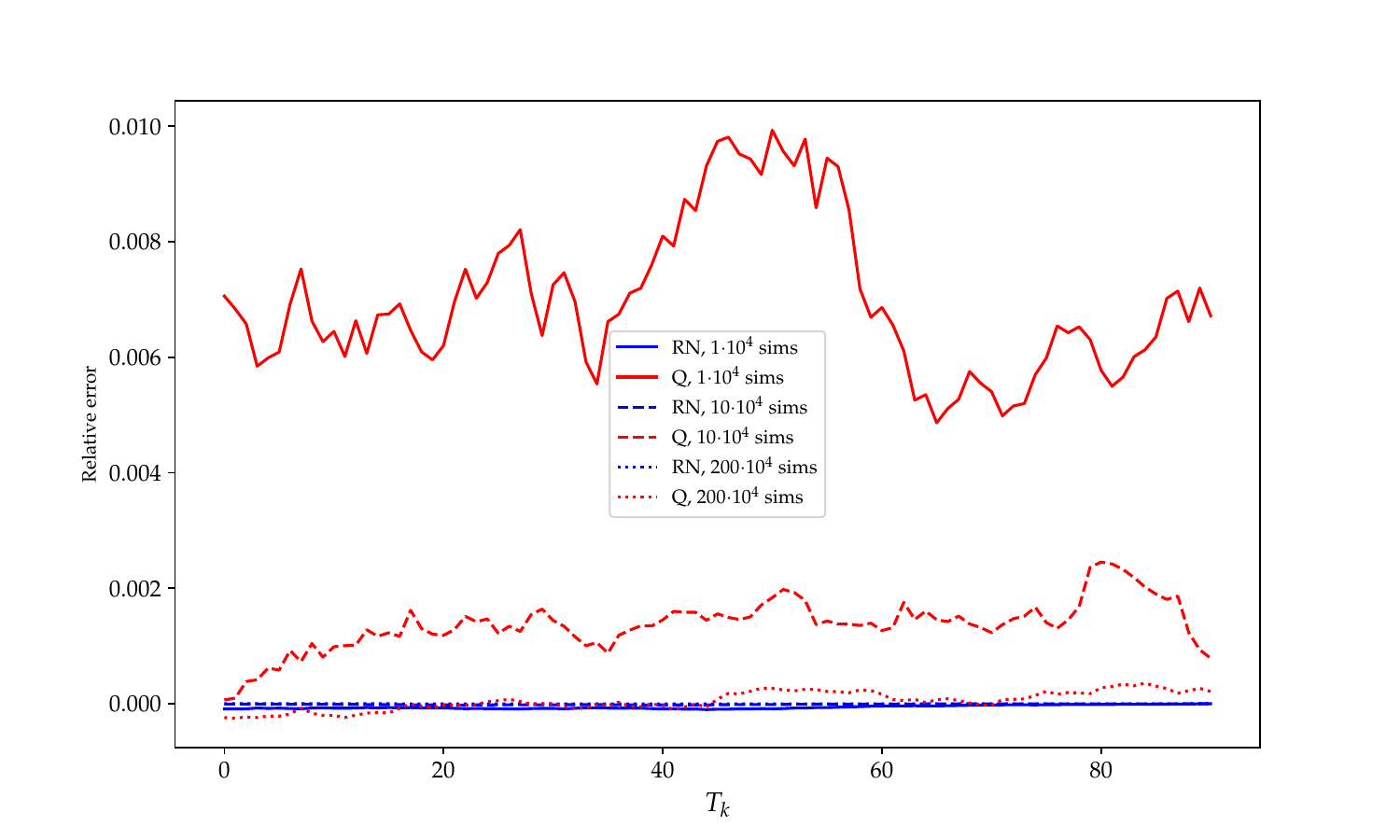}
		\caption{Case 1.}
		\label{fig:relative_errors_case_1}
	\end{subfigure}
	\hfill
	\begin{subfigure}[b]{0.9\textwidth}
		\centering
		\includegraphics[width=\linewidth]{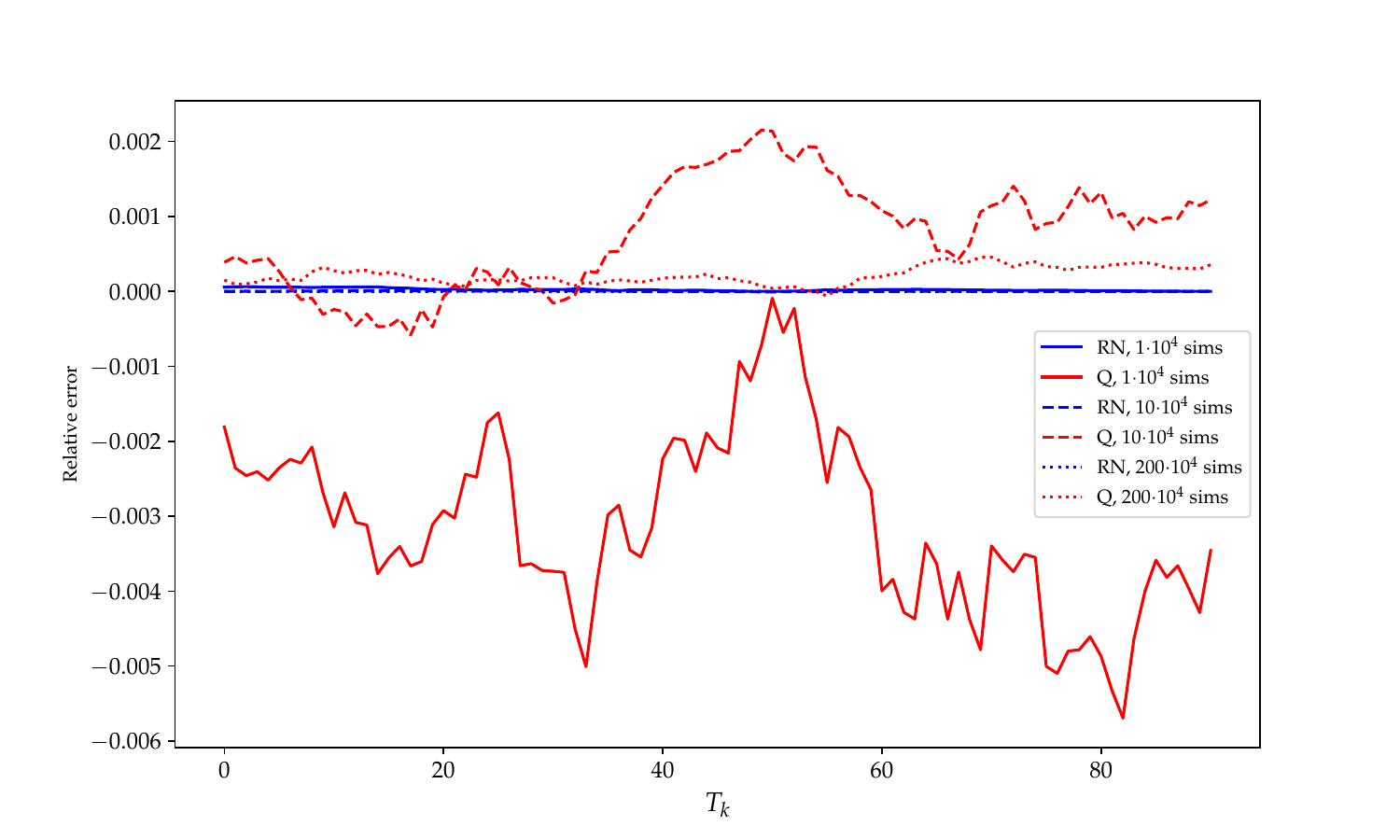}
		\caption{Case 2.}
		\label{fig:relative_errors_case_2}
	\end{subfigure}
	\caption{Relative errors for the different expressions for the arithmetic factors. The number of simulations is varied between $10^4$ and $2\cdot 10^6$. The parameters are $T_s=1$ month and $T_e-T_s=3$ months. The cases are computed using the same set of parameters, but different draws of the Brownian motion.}
	\label{fig:relative_errors}
\end{figure}

 A similar conclusion can be drawn from the relative errors, see Figure \ref{fig:relative_errors}. There, we have computed the relative errors for the different expressions for the arithmetic factors, varying the number of simulations between $10^4$ and $2\cdot 10^6$. We have computed the relative errors for two cases, that is, with different seeds for the Brownian motion. As we can see, the relative errors are smaller for the case with the product of two normalized variables and quite ``unstable'' for the case with the arithmetic factor computed using the quotient of expectations form.
\clearpage
\section{Theoretical bounds and closed formula for the Gaussian HJM model}
\label{sec:GHJM}
As we can see from the plots, it seems to be the case that $\mathcal{A}_1 \le 1$. Let us prove this for one-factor model (that is, Hull-White or Hagan's LGM model).

\begin{proposition}\label{prop:Ak_le_1}
	Assume the short rate follows \eqref{eq:sr model} with $n=1$, then 
	$$\mathcal{A}_k \le 1\quad\text{for all }k \in \{1, \ldots, K\}\,.$$ 
\end{proposition}
\begin{proof}
	To simplify the proof, let us work in the HJM framework\footnote{We will explore more on this, although using a slightly different approach, below.}, following Andersen and Piterbarg's book. It is easier if we work with the expression,
	$$
	\mathcal{A}_{k} = \frac{\mathbb{E}^{T_e} \left( R_k \right)}{\mathbb{E}^{T_k} \left( R_k \right)}\,.
	$$
	By \cite[Remark 10.1.8]{AP10b}, the discount bond dynamics for $P(t, T)$ are given by
	$$
	d P(t, T) / P(t, T) = r(t) d t - \sigma_P(t, T) d W(t), \quad \sigma_P(t, T) = \sigma_r(t) G(t, T)\,,
	$$
	with $G(t, T) \coloneqq \int_t^T e^{-\int_t^u \varkappa(s) d s} d u$, $\sigma_r$ the short rate volatility (now a deterministic function), and $\varkappa$ the mean reversion speed. That is, if $x(t) \coloneqq r(t) - f(0,t)$ by \cite[Proposition 10.1.7]{AP10b},
	$$
	d x(t) = (y(t) - \varkappa(t) r(t)) d t + \sigma_r(t) d W(t)\,.
	$$	
	Thus, as the SDE with deterministic functions as coefficients,
	$$
	d x(t) = (A(t) x(t) + \tilde{B}(t)) d t + C(t) d W(t)\,,
	$$
	admits the unique solution \cite[(8.2.5) Corollary]{Arn74} given by the fundamental ``matrix'',
$$
\Phi(t)=\exp \left(\int_{0}^t A(s)      	 {d} s\right)\,,
$$
i.e.,
$$
x(t)=\exp \left(\int_{0}^t A(s) {d} s\right)\left(x(0)+\int_{0}^t \exp \left(-\int_{0}^s A(u) {d} u\right)\left(\tilde{B}(s) {d} s+C(s) {d} W_s\right)\right) .
$$
	we can solve our equation. Using \cite[Equation (4.34)]{AP10a}, Girsanov's Theorem reads as
	\begin{equation*}
		d W^T(t) = d W(t) + \sigma_P(t, T) d t\,,
	\end{equation*}
	where $W^T$ is a $Q^T$-Brownian motion. This introduces a new term in the drift, in particular in $\tilde{B}$, of the form $-\sigma_r^2(t) G(t, T)$. For $T' \ge T$, $\sigma_r^2(t) G(t, T') \ge \sigma_r^2(t) G(t, T)$. Thus, if $Z \mid_Q$ denotes the distribution of any random variable $Z$ under the measure $Q$, and $X^{T'}_t \sim x(t) \mid_{Q^{T'}}$, then
	$$
	X^T_t := X^{T'}_t + \beta(t) \sim x(t) \mid_{Q^T}\,,
	$$ 
	where 
	$$\beta(t)\coloneqq \exp \left(\int_{0}^t -\kappa(s) {d} s\right)\left(\int_{0}^t \exp \left(\int_{0}^s \kappa(u) {d} u\right)\left(\sigma_r^2(s) G(s, T')-\sigma_r^2(s) G(s, T)\right) {d} s\right) \ge 0$$
	is a deterministic function. As \cite[Proposition 10.1.7]{AP10b},
	$$
	P(t, T, x(t)) = \frac{P(0, T)}{P(0, t)} \exp \left(-x(t) G(t, T) - \frac{1}{2} y(t) G(t, T)^2\right)\,,
	$$
	thus\footnote{Using the aforementioned law of the unconscious statistician.},
	\begin{equation*}
		\mathbb{E}^{T_k} \left(\frac{1}{P(T_{k-1}, T_{k}, x(T_{k-1}))}\right) = \mathbb{E}^{T_e} \left(\frac{1}{P(T_{k-1}, T_{k}, x(T_{k-1}) + \beta(T_k))}\right) \ge \mathbb{E}^{T_e} \left(\frac{1}{P(T_{k-1}, T_{k}, x(T_{k-1}))}\right)\,,
	\end{equation*}
	as $G(T_{k-1}, T_{k}) > 0.$
\end{proof}
\begin{remark}\label{rem:Ak_leq_1}
	It is reasonable to assume that we are going to have \( A_k \leq 1 \), but not necessarily for \( \mathcal{A}_k \). Indeed, we have:
	\begin{align*}
		\mathbb{E}^Q \left( e^{-\int_0^{T_k} r(s) \, ds} R_k \right) &= P(0, T_k) \cdot \mathbb{E}^{T_k} (R_k) = P(0, T_k) \cdot F_k\,.
	\end{align*}
	By the law of iterated expectations,
	\begin{align*}
		\mathbb{E}^Q \left( e^{-\int_0^{T_e} r(s) \, ds} R_k \right) &= \mathbb{E}^{Q} \left( e^{-\int_0^{T_k} r(s) \, ds} \, P(T_k, T_e) R_k \right) \\
		&\leq \mathbb{E}^Q \left( e^{-\int_0^{T_k} r(s) \, ds} R_k \right) = P(0, T_k) \cdot F_k\,,
	\end{align*}
	where we have assumed that \( P(T_k, T_e) \leq 1 \) and $R_k\geq 0$ almost surely, a sufficient condition being $r(t)$ non-negative. Note that for the equality to hold, it is sufficient that the measure of the subset where \( P(T_k, T_e) R_k > R_k \) is small enough\footnote{Indeed, let $\Omega_0$ be the subset where $P(T_k, T_e) R_k < R_k$ and $\Omega_0^c$ the subset where $P(T_k, T_e) R_k \geq R_k$. Then,
	$$
	\mathbb{E}^Q \left( e^{-\int_0^{T_k} r(s) \, ds} \, (P(T_k, T_e) - 1) R_k \right) = \mathbb{E}^Q \left( \mathbb{1}_{\Omega_0} e^{-\int_0^{T_k} r(s) \, ds} \, (P(T_k, T_e) - 1) R_k \right) + \mathbb{E}^Q \left( \mathbb{1}_{\Omega_0^c} e^{-\int_0^{T_k} r(s) \, ds} \, (P(T_k, T_e) - 1) R_k \right)\,,
	$$
	but if $ H_k\coloneqq \exp\left({-\int_{0}^{T_k} r}\right) \left( P(T_k, T_e) - 1\right)R_k$ is in $L^p(\Omega)$ for some $p>1$, then the integral satisfies
	\[
		\mathbb{E}^Q \left[ \mathbb{1}_{\Omega_0^c}  H_k \right] \le  \mathbb{E}^Q \left[ 
		 H_k ^p \right] ^{1/p}  Q(\Omega_0^c) ^{1/q} \to 0 \text{ as } \, Q(\Omega_0^c) \to 0\,.
	\]
	where $1/p+1/q=1$ and we have used H\"older's inequality. Alternatively, the Dominated Convergence Theorem can be used to show that the limit goes to zero in the case of $p=1$ taking a decreasing sequence of $\Omega_0^c$. Note that the fact of the measure being small enough depends on the equivalent measure used.}, depending on the values of \( R_k \) and the discount on that region. Therefore, \( A_k \leq 1 \). However, the sign of \( \mathcal{A}_k - 1 \) is undetermined a priori since
	\[
	\mathcal{A}_k = A_k \cdot \frac{P(0, T_k)}{P(0, T_e)}\, \text{ with, generally, } \frac{P(0, T_k)}{P(0, T_e)} \geq 1\,.
	\]
\end{remark}
\subsection{Gaussian HJM models}
Following \cite[Chapter 12]{AP10b} (see also Chapter 22 of \cite{Bjo20} or Appendix A of \cite{Hen14}), let us consider a general \textit{d}-factor Gaussian model that can be written as \cite[(4.31)]{AP10a},
\begin{equation}\label{eq:bond dynamics}
	\frac{dP(t,T)}{P(t,T)} = r(t) dt - \sigma_P(t,T)^\top d\tilde W(t),
\end{equation}
where $\sigma_P(t,T)$ is a bounded \textit{n}-dimensional function of time, and $\tilde W(t)$ a \textit{n}-dimensional (uncorrelated) Brownian motion in the risk-neutral measure ${Q}$. Then, the instantaneous forward rates, $f(t,T)=:f^T(t)$, are given by \cite[(4.33), the HJM condition]{AP10a},
\begin{equation*}
	df(t,T) := df^T(t) = \sigma_f(t,T)^\top \sigma_P(t,T) dt + \sigma_f(t,T)^\top d\tilde W(t)
\end{equation*}
\begin{equation}
	= \sigma_f(t,T)^\top \int_t^T \sigma_f(t,u) du \, dt + \sigma_f(t,T)^\top d\tilde W(t).
\end{equation}
This model is generally not Markovian \cite[p.184]{AP10a}, so we impose the following condition.

\begin{assumption}[Separability condition]\label{as:sep cond}
	Assume that $\sigma_f(t,T)$ is \textit{separable}, i.e.,
	\begin{equation}
		\sigma_f(t,T) = g(t) h(T), 
	\end{equation}
	where $g$ is a $n \times n$ deterministic matrix-valued function, and $h$ is a \textit{n}-dimensional deterministic vector. 
\end{assumption}

The full theory for these models is developed in Chapter 12 of \cite{AP10b}, but from here we can compute explicitly $\mathcal{A}_k$. See also \cite[Theorem 6.4]{Hen14} for a different proof of this result \footnote{Some minor typos are present in that formula.}.
\begin{proposition}
	Assume \eqref{eq:bond dynamics} holds. Then,
	\begin{equation*}
		\mathcal{A}_k = \gamma_k + \frac{\gamma_k - 1}{\tau_k F_k}\,,
	\end{equation*}
	where the convexity adjustment for deferred payment is given by
	\begin{equation*}
		\gamma_k = \exp\left( \int_0^{T_{k-1}} (\sigma_P(s, T_{k-1}) - \sigma_P(s, T_{k}))^\top \cdot (\sigma_P(s, T_e) - \sigma_P(s, T_{k})) \, ds \right)\,.
	\end{equation*}
\end{proposition}
Note that if $T_e=T_k$, i.e., $k=K$, then $\gamma_{K}=1$ and $\mathcal{A}_{K}=1$, as expected.
\begin{proof}
For the sake of simplicity in our computations, let us define:
\begin{equation*}
	\sigma_k \coloneqq \sigma_P(s, T_k)^\top\,, \quad \sigma_e \coloneqq \sigma_P(s, T_e)^\top\,, \quad \sigma_k^2\coloneqq \sigma_k \cdot \sigma_k^\top \,.
\end{equation*}
It is standard that, see for instance Appendix A of \cite{Hen14},
\begin{equation*}
	P(T_{k-1}, T_{k}) = P(0, T_{k-1}, T_{k}) \exp \left( -\int_0^{T_{k-1}} (\sigma_{k} - \sigma_{k-1}) \, dW_s - \frac{1}{2} \int_0^{T_{k-1}} (\sigma_{k}^2 - \sigma_{k-1}^2) \, ds \right)\,,
\end{equation*}
being  $P^M(0; T_{k-1}, T_{k})$ the forward bond price at time $0$, so obtained from market prices, for the period between $T_{k-1}$ and $T_{k}$. Now, considering the conditional expectation under the measure $\mathbb{E}^{T_e}$:
\begin{align*}
	\mathbb{E}^{T_e}({R_k}) &= \mathbb{E}^{T_e} \left( \frac{1}{\tau_k} \left( \frac{1}{P(T_{k-1}, T_k)} - 1 \right) \right) \nonumber \\
	&= \frac{1}{\tau_k} \cdot \frac{1}{P^M(0, T_{k-1}, T_{k})} \, \mathbb{E}^{T_e} \left(  \exp \left( \int_0^{T_{k-1}} (\sigma_{k} - \sigma_{k-1}) \, dW_s^{T_e} \right) \right. \nonumber \\
	&\quad \times \left. \exp \left( -\frac12\int_0^{T_{k-1}} (\sigma_{k}^2 - \sigma_{k-1}^2) \, ds + \int_0^{T_{k-1}} (\sigma_{k} - \sigma_{k-1}) \sigma_e^\top \, ds \right) \right)-\frac{1}{\tau_k}.
\end{align*}
where we have used the following Girsanov's transformation, see \cite[(4.34)]{AP10a},
\begin{equation}
	dW_s = dW^{T_e}_s + \sigma_e \, ds\,.
\end{equation}
As $\sigma$ is deterministic, from the moment generating function of a normal variable and Itô's isometry, we obtain
\begin{equation*}
	\mathbb{E}^{T_e} \left(  \exp \left( \int_0^{T_{k-1}} (\sigma_{k} - \sigma_{k-1}) \, dW_s^{T_e} \right)\right) = \exp \left(\frac12 \int_0^{T_{k-1}} (\sigma_{k} - \sigma_{k-1})^2 \, ds \right)\,.
\end{equation*}
Therefore:
\begin{align*}
	\mathbb{E}^{T_e}(R_k) &= \frac{1}{\tau_k} \cdot \frac{1}{P^M(0, T_{k-1}, T_{k})} \exp\left( \int_0^{T_{k-1}} \sigma_{k}^2 - \sigma_{k-1}\cdot \sigma_{k}^T - (\sigma_{k} - \sigma_{k-1}) \sigma_e^\top\, ds \right) - \frac{1}{\tau_k} \\
	&= \frac{1}{\tau_k} \cdot \frac{1}{P^M(0, T_{k-1}, T_{k})} \gamma_k - \frac{1}{\tau_k}, 
\end{align*}
where the convexity adjustment for deferred payment is given by
\begin{align*}
	\gamma_k &\coloneqq \exp\left( -\int_0^{T_{k-1}} (\sigma_{k} - \sigma_{k-1})(\sigma_e - \sigma_{k})^\top \, ds \right)\\
	&=\exp\left(- \int_0^{T_{k-1}} (\sigma_P(s, T_{k}) - \sigma_P(s, T_{k-1}))^\top \cdot (\sigma_P(s, T_e) - \sigma_P(s, T_{k})) \, ds \right)\,.
\end{align*}
Finally, we have:
\begin{align*}
	\mathcal{A}_k &= \frac{\mathbb{E}^{T_e}(R_k)}{\mathbb{E}^{T_k}(R_k)} = \left( \frac{F_k + 1 / \tau_k}{F_k} \right) \gamma_k-\frac{1}{\tau_k F_k}= \gamma_k + \frac{\gamma_k - 1}{\tau_k F_k}\,. 
\end{align*}
\end{proof}
\begin{remark}
	Let us explore the sign of $\mathcal{A}_k - 1$; see Proposition~\ref{prop:Ak_le_1}. It is useful to work with the LGM (Linear Gaussian Markov) with $n$ factors; see \cite{Rom24b} for details. That is,
	\[
	H(T) = \int_0^T h(s)^\top ds,\quad g(t) = C^\top A(t),
	\]
	where $\rho = C C^\top$ is the correlation matrix of $W$, and $A_{ij}(t) = \alpha_i(t) \delta_{ij}$. Here, $dZ_i(t) = \alpha_i(t)\, dW_i^N(t)$ is the $i$-th factor in the measure $Q^N$; see the aforementioned reference for more details. Then, the convexity adjustment reads as
	\begin{equation}
		\log \gamma_k = -\left( H(T_{k}) - H(T_{k-1}) \right) \zeta(T_{k-1}) \left( H(T_e) - H(T_{k}) \right)^\top,
	\end{equation}
	with $\zeta(T_k)$ being the covariance matrix of the factors at time $T_k$. It is standard to assume that the components of $h(s)$ are non-negative and non-increasing, so we have a mean-reverting short rate rather than a mean-fleeing one; see Equation (12.8) in \cite{AP10b}. Therefore, the components of $H(\cdot)$ are non-decreasing since its derivative is non-negative.
	In the case of $n = 1$, this is enough to ensure that $\gamma_k \leq 1$, but this is not sufficient when $n > 1$, as one can easily construct two componentwise positive vectors which, when multiplied through a positive definite matrix, return a negative value.
\end{remark}
\subsubsection{Example: the Hull-White model with constant coefficients} \label{sec:HW_case}
\begin{figure}[h!]
	\centering
	\begin{subfigure}[b]{0.7\textwidth}
		\centering
		\includegraphics[width=\linewidth]{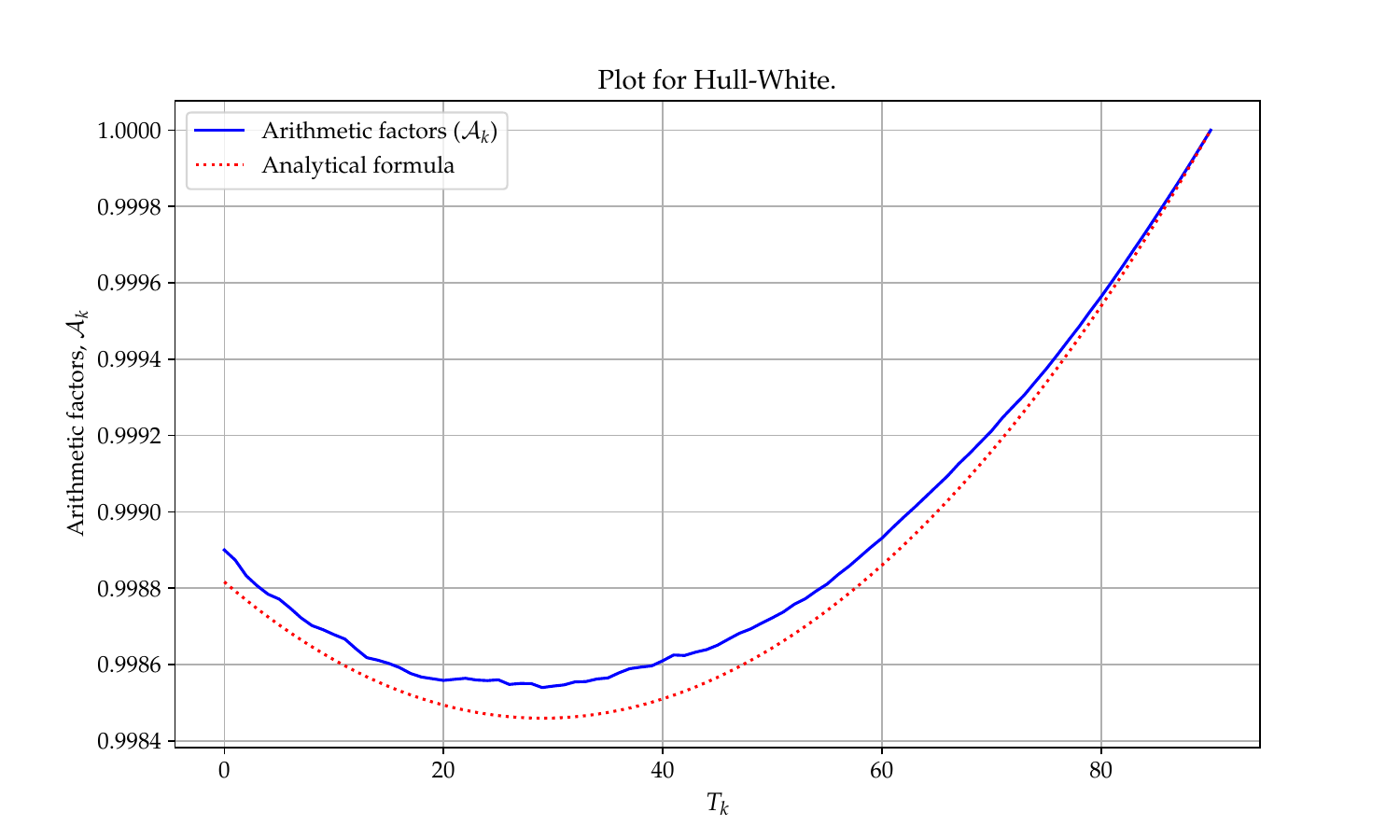}
		\caption{$10^4$ simulations.}
		\label{fig:HW_A_k_10e3}
	\end{subfigure}
	\hfill
	\begin{subfigure}[b]{0.7\textwidth}
		\centering
		\includegraphics[width=\linewidth]{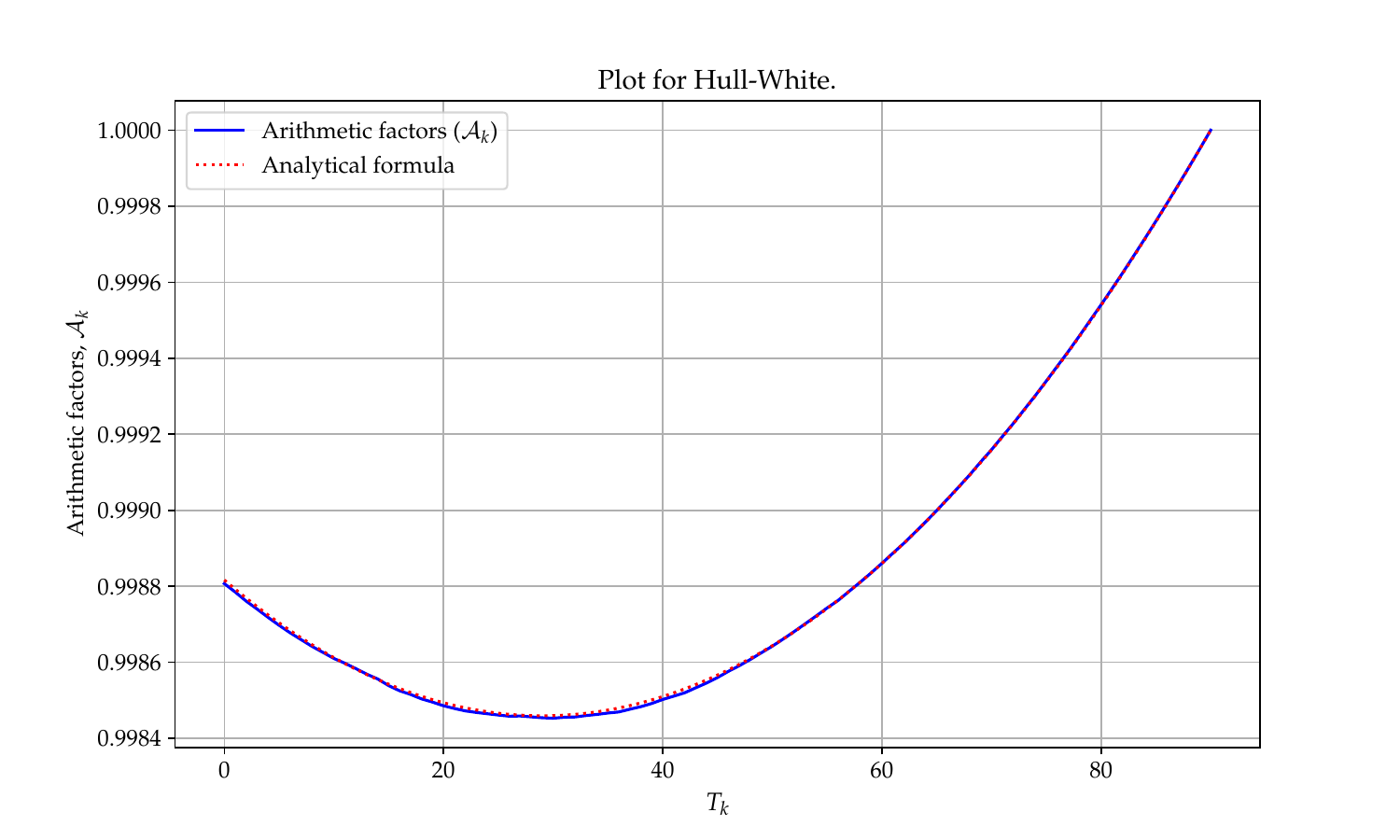}
		\caption{$10^5$ simulations.}
		\label{fig:HW_A_k_100e3}
	\end{subfigure}
	\hfill
	\begin{subfigure}[b]{0.7\textwidth}
		\centering
		\includegraphics[width=\linewidth]{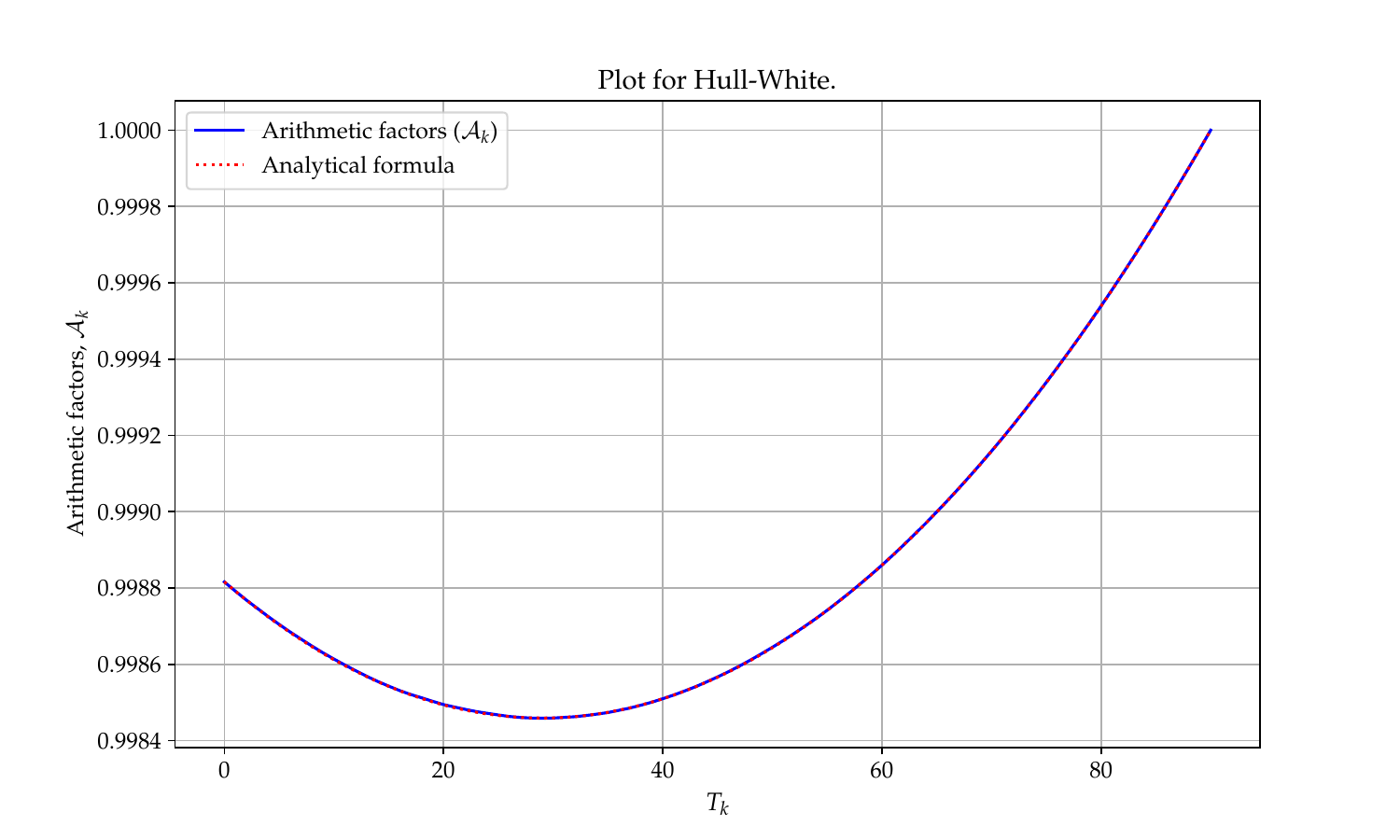}
		\caption{$2 \times 10^6$ simulations.}
		\label{fig:HW_A_k_2000e3}
	\end{subfigure}
	\caption{Comparison between the analytical expression for \(\mathcal{A}_k\) using the above expression for \(\gamma_k\) and the numerical approximation for different number of simulations. Here, $T_s=1$ month and $T_e-T_s=3$ months.}
	\label{fig:HW_A_k}
\end{figure}

Given the Hull-White model with volatility function, see Remark 10.1.8 of \cite{AP10b}, \(\sigma_P(s, t) = \frac{\eta}{a} \left( 1 - e^{-a(t - s)} \right)\), where both \(\eta\) and \(a\) are constants, we can develop the particular expressions for this case.
First, recall the definition of \(\mathcal{A}_k\):
\begin{equation*}
	\mathcal{A}_k = \gamma_k + \frac{\gamma_k - 1}{\tau_k F_k}\,,
\end{equation*}
where the convexity adjustment \(\gamma_k\) for deferred payment is given by
\begin{equation*}
	\gamma_k = \exp\left( \int_0^{T_{k-1}} \left[ \sigma_P(s, T_{k-1}) - \sigma_P(s, T_{k}) \right] \left[ \sigma_P(s, T_e) - \sigma_P(s, T_{k}) \right] \, ds \right)\,.
\end{equation*}
Substituting the given \(\sigma_P(s, t)\) into the expression for \(\gamma_k\), we compute the integrand:
\begin{align*}
	\sigma_P(s, T_{k-1}) - \sigma_P(s, T_{k}) &= \frac{\eta}{a} \left( e^{-a(T_{k} - s)} - e^{-a(T_{k-1} - s)} \right)\,, \\
	\sigma_P(s, T_e) - \sigma_P(s, T_{k}) &= \frac{\eta}{a} \left( e^{-a(T_{k} - s)} - e^{-a(T_e - s)} \right)\,.
\end{align*}
The product of these differences is:
\begin{align*}
	\left[ \sigma_P(s, T_{k-1}) - \sigma_P(s, T_{k}) \right]&\left[ \sigma_P(s, T_e) - \sigma_P(s, T_{k}) \right] \\
	&= \left( \frac{\eta}{a} \right)^2 \left( e^{-a(T_{k} - s)} - e^{-a(T_{k-1} - s)} \right) \left( e^{-a(T_{k} - s)} - e^{-a(T_e - s)} \right)\,.
\end{align*}
Expanding the product:
\begin{align*}
	&\left( e^{-a(T_{k} - s)} - e^{-a(T_{k-1} - s)} \right) \left( e^{-a(T_{k} - s)} - e^{-a(T_e - s)} \right) \\
	&= e^{-a(T_{k} + T_k - 2s)} - e^{-a(T_{k} + T_e - 2s)} - e^{-a(T_{k-1} + T_{k} - s)} + e^{-a(T_{k-1} + T_e - 2s)}\,.
\end{align*}
Integrate this expression with respect to \(s\) from \(0\) to \(T_e\):
\begin{equation*}
	\int_0^{T_{k-1}} \left( \cdots \right) ds = \frac{1}{2a} \left( e^{2a T_{k-1}} - 1 \right) \left[ e^{-2aT_{k}} - e^{-a(T_{k} + T_e)} - e^{-a( T_k+T_{k-1})} + e^{-a(T_{k-1} + T_e)} \right]\,.
\end{equation*}
Thus, the convexity adjustment \(\gamma_k\) simplifies to:
\begin{equation*}
	\gamma_k = \exp\left( \left( \frac{\eta}{a} \right)^2 \frac{e^{2a T_{k-1}} - 1}{2a} \left[ e^{-2aT_{k}} - e^{-a(T_{k} + T_e)} - e^{-a( T_k+T_{k-1})} + e^{-a(T_{k-1} + T_e)} \right] \right)\,.
\end{equation*}
A comparison between the analytical expression for \(\mathcal{A}_k\) using the above expression for \(\gamma_k\) and the numerical approximation was shown in Figure \ref{fig:comparison} and Figure \ref{fig:relative_errors}, where a close alignment is observed. In order to appreciate the accuracy of the approximation, we also show another plot in Figure \ref{fig:HW_A_k}, where now the factors are much closer to one, so absolute errors are easier to appreciate. See also Remark~\ref{rem:U_shape}.
\clearpage
\section{Connection with Takada's approximation}

In this section, we explore the relationship between the approximation developed in the previous sections and an approximation suggested by Katsumi Takada in \cite{Tak11}. We start with the following trivial expression:
	
\begin{equation}
	\sum \tau_k \cdot F_k = \log \left( e^{\sum \tau_k F_k} \right)\,.
\end{equation}
Given that \( e^{x} = 1 + x +  o(x) \) for small \( x \). In this case, $o(x)>0$ for positive values of $x$. Therefore, using the definition of little-o,
$$
e^{\sum_{k=1}^K \tau_k F_k} = \prod_{k=1}^K e^{\tau_k F_k}=\prod_{k=1}^K\left(1+\tau_k F_k+o\left(\tau_k F_k\right)\right)=\prod_{k=1}^K\left(1+\tau_k F_k\right)+\sum_{k=1}^{K}o\left(\mu^k\right)\,,
$$
being $\mu\coloneqq \max_k{|\tau_k F_k|}$. Thus,
$$
e^{\sum_{k=1}^K \tau_k F_k} = \prod_{k=1}^K\left(1+\tau_k F_k\right)+o\left(\mu\right)\,,
$$
being the error term strictly positive. Expanding the leading term, this is equal to:
\begin{align*}
\log \left( \prod_{k=1}^{K} (1 + \tau_k F_k) \right)&=	\log \left( \prod_{k=1}^{K} \frac{P( T_{k-1})}{P( T_{k })} \right)\\
&=	\log \left( \frac{P( T_s)}{P( T_e)} \right)\equiv \log \left( 1 + \tau(T_s, T_e) F_g (0; T_s, T_e) \right)\,,
\end{align*}
where the last equality follows by the definition of continuously compounded or geometric forwards and a telescopic cancellation for the product. Therefore, we obtain, cf. (7) and (8) of \cite{Tak11} or (145) of  \cite{AB13},
\begin{equation}
\sum \tau_k \cdot F_k \approx	\log \left( 1 + \tau(T_s, T_e) F_g (0; T_s, T_e) \right)\eqqcolon\tau(T_s, T_e)\cdot O^\text{det}_a(0;T_s,T_e)\,,
\end{equation}
following (13) of \cite{Tak11}, a deterministic version of Takada's forward definition, which he denotes by $O_a(0; T_s,T_e)$. That is, 
$$
	{O}_a(0;T_s,T_e) \coloneqq \frac{1}{\tau(T_s,T_e)}\mathbb{E}^{T_e} \left( \int_{T_s}^{T_e} r \right)\,
$$
and for deterministic rates $O^\text{det}_a(0;T_s,T_e)=O_a(0;T_s,T_e)$. 
Thus, if $\mathcal{A}_k\approx 1$, the forward rate approximation \( F_a(0, T_s, T_e) \) is given by:

\begin{equation}\label{eq:Tak fin}
	F_a(0, T_s, T_e) \approx O^\text{det}_a(0;T_s,T_e) > O_a (0, T_s, T_e)\,,
\end{equation}
by (15) of \cite{Tak11}.
\begin{remark}
	 Note that (15) of \cite{Tak11} or the last inequality of \eqref{eq:Tak fin} is proven using a no-arbitrage argument. Nevertheless, a ``pure '' mathematical argument can be given too. Indeed, using Jensen's inequality, assuming that the integral is strictly positive,
	\begin{align*}
		\delta(T_s,T_e){O}_a(0;T_s,T_e) = \mathbb{E}^{T_e} \left( \int_{T_s}^{T_e} r \right) &\overset{\text{Jensen's ineq.}}{<} \log \mathbb{E}^{T_e} \left( e^{\int_{T_s}^{T_e} r} \right)\\& = \log \left( \frac{1}{P(T_e)} \cdot \mathbb{E}^Q \left( e^{-\int_{0}^{T_s} r} \right) \right)\,,
	\end{align*}
	where we have used that for any derivative with value $V$, the change of numeraire formula reads as,
	$$
	P(t, T_e) \mathbb{E}^{T_e} \left( {V_{T_e}} \right) = V_t = B(t)\cdot \mathbb{E}^Q \left(  e^{-\int_{0}^{T_e}r}{V_{T_e}} \right)\,,
	$$
	Thus, we conclude the proof noting that $\mathbb{E}^Q \left( e^{-\int_{0}^{T_s} r} \right)=P(T_s)$.

\end{remark}

\printbibliography

\end{document}

%% file: tables/results_tenor_3M_delay_1M_no_sims_100e3_models_False.tex
\begin{tabular}{@{}l|c|c|c|c|c@{}}
$(\sigma, a, \eta, b, \rho)$ & $\mathcal{A}_1$ & $F_a$ & $\frac{F_a^\text{Mx}}{F_a}-1$ & $\frac{F_a^\text{lin}}{F_a}-1$ & $\frac{F_a^\text{pw}}{F_a}-1$ \\
\hline
(0.03, 0.29, 0.05, 0.78, -0.22) & 0.99893 & 0.04995 & 0.00101 & 0.00047 & 0.00012 \\
(0.04, 0.97, 0.03, 0.24, 0.17) & 0.99882 & 0.04995 & 0.00111 & 0.00052 & 0.00014 \\
(0.07, 0.72, 0.03, 0.97, 0.22) & 0.99751 & 0.04989 & 0.00237 & 0.00112 & 0.00029 \\
(0.02, 0.03, 0.01, 0.75, 0.07) & 0.99979 & 0.04999 & 0.00020 & 0.00010 & 0.00003 \\
(0.03, 0.17, 0.08, 0.31, -0.28) & 0.99740 & 0.04988 & 0.00251 & 0.00121 & 0.00032 \\
\end{tabular}

%% file: tables/results_tenor_6M_delay_12M_no_sims_100e3_models_False.tex
\begin{tabular}{@{}l|c|c|c|c|c@{}}
$(\sigma, a, \eta, b, \rho)$ & $\mathcal{A}_1$ & $F_a$ & $\frac{F_a^\text{Mx}}{F_a}-1$ & $\frac{F_a^\text{lin}}{F_a}-1$ & $\frac{F_a^\text{pw}}{F_a}-1$ \\
\hline
(0.06, 0.01, 0.07, 0.42, -0.28) & 0.95344 & 0.04866 & 0.02758 & 0.00366 & 0.00104 \\
(0.02, 0.1, 0.08, 0.58, -0.61) & 0.97530 & 0.04931 & 0.01414 & 0.00162 & 0.00047 \\
(0.09, 0.97, 0.04, 0.45, 0.25) & 0.95313 & 0.04869 & 0.02708 & 0.00301 & 0.00087 \\
(0.08, 0.49, 0.05, 0.96, -0.73) & 0.97926 & 0.04941 & 0.01196 & 0.00147 & 0.00044 \\
(0.09, 0.71, 0.06, 0.26, 0.63) & 0.89445 & 0.04702 & 0.06350 & 0.00738 & 0.00217 \\
\end{tabular}